\documentclass[a4paper]{jpconf}
\usepackage{graphicx}

\usepackage[symbol]{footmisc}
\renewcommand{\thefootnote}{\fnsymbol{footnote}}

\begin{document}
\title{Current Status and New Challenges of The Tunka Radio Extension}

\author{V. Lenok$^1$,
P.A. Bezyazeekov$^2$,
N.M. Budnev$^2$,
D. Chernykh$^2$,
O. Fedorov$^2$,
O.A. Gress$^2$,
A. Haungs$^1$,
R. Hiller$^1$\footnote{now at the University of Z\"urich},
T. Huege$^1$\footnote{also at Vrije Universiteit Brussel, Belgium},
Y. Kazarina$^2$,
M. Kleifges$^3$,
D. Kostunin$^1$,
E.E. Korosteleva$^4$,
L.A. Kuzmichev$^4$,
N. Lubsandorzhiev$^4$,
T. Marshalkina$^2$,
R. Monkhoev$^2$,
E. Osipova$^4$,
A. Pakhorukov$^2$,
L. Pankov$^2$,
V.V. Prosin$^4$,
F.G. Schr\"oder$^1,^5$,
D. Shipilov$^2$ and
A. Zagorodnikov$^2$
(Tunka-Rex Collaboration)
}

\address{$^1$ Institut f\"ur Kernphysik, Karlsruhe Institute of Technology (KIT), Karlsruhe, Germany}
\address{$^2$ Institute of Applied Physics ISU, Irkutsk, 664020, Russia}
\address{$^3$ Institut f\"ur Prozessdatenverarbeitung und Elektronik, Karlsruhe Institute of Technology (KIT)}
\address{$^4$ Skobeltsyn Institut of Nuclear Physics MSU, Moscow, 119991, Russia}
\address{$^5$ Department of Physics and Astronomy, University of Delawere, Newark, DE, USA}

\ead{vladimir.lenok@kit.edu}

\begin{abstract}
  The Tunka Radio Extension (Tunka-Rex) is an antenna array spread over an area of about 1~km$^2$.
  The array is placed at the Tunka Advanced Instrument for cosmic rays and Gamma Astronomy (TAIGA)
  and detects the radio emission of air showers in the band of 30 to 80~MHz.
  During the last years it was shown that a sparse array such as Tunka-Rex is capable of reconstructing
  the parameters of the primary particle as accurate as the modern instruments.
  Based on these results we continue developing our data analysis.
  Our next goal is the reconstruction of cosmic-ray energy spectrum observed only by a radio instrument.
  Taking a step towards it, we develop a model of aperture of our instrument
  and test it against hybrid TAIGA observations and Monte-Carlo simulations.
  In the present work we give an overview of the current status and results for the last five years of operation
  of Tunka-Rex and discuss prospects of the cosmic-ray energy estimation with sparse radio arrays.
\end{abstract}

\section{Introduction}

Digital radio antenna arrays are a rapidly developing instrument for observation and
study of ultra-high energy messengers in PeV-EeV range~\cite{schroeder2017,huege2016}.
Radio detection has evolved from understanding signal formation and early hardware development
to the stage of real physical measurements.

\begin{figure}[h]
  \begin{center}
    \includegraphics[width=1.\textwidth]{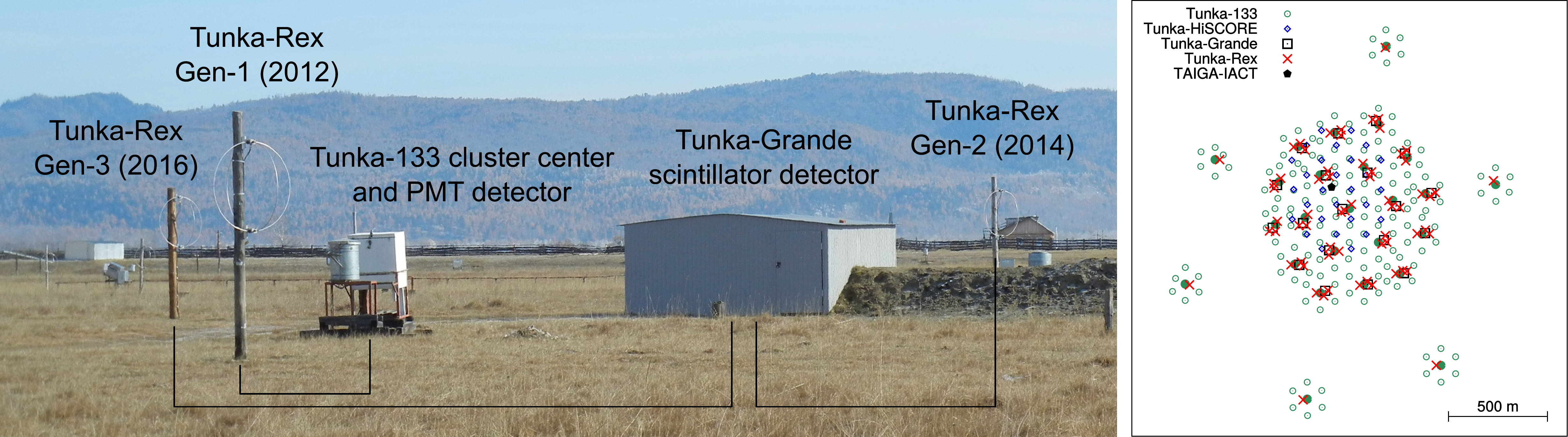}
  \end{center}
  \caption{\label{cluster}
  {\it Left}:
  one of the clusters of Tunka-Rex.
  Three generations of the antennas are marked in the picture along with the Tunka-Grande and Tunka-133 detectors. The black lines schematically represent the cabling.
  {\it Right}:
  map of the present footprint of the detectors of TAIGA.
  }
\end{figure}

Tunka-Rex is a radio antenna array designed for detecting cosmic rays.
It is located at the site of the TAIGA detector~\cite{taiga}
at the Tunka valley in Russia.
The Tunka-Rex detector has been operating since 2012.
It has been extended several times and
currently consists of 57 antennas in the dense core of the detector (about 1~km$^2$ area) and
6 antennas in the satellite clusters used as a high-energy extension.
In total the Tunka-Rex radio array covers about 3~km$^2$.
As antenna type we use SALLA~\cite{salla} operating in the band of 30--80~MHz.
The present configuration of Tunka-Rex detector is shown in the Fig.~\ref{cluster}.

\section{Highlights of Recent Results}

Some important updates of the analysis have been done in Tunka-Rex.
Recently we developed a method of template analysis of the data.
Essentials of this method are the following:
\begin{itemize}
  \item Pre-reconstruction using the standard procedure of Tunka-Rex~\cite{jcap}.
    This stage provides the shower incoming direction, the core position, and the energy estimation.
  \item Creating a library of CoREAS simulations~\cite{coreas} for each event obtained in the previous step
    with the goal to cover all possible depths of the shower maxima.
  \item Least-square fit of the simulated signal envelops to the reconstructed ones.
    The shower maximum and primary-particle energy are reconstructed on this stage.
\end{itemize}

\begin{figure}[h]
\begin{center}
\begin{minipage}{17pc}
\begin{center}
\includegraphics[height=13pc]{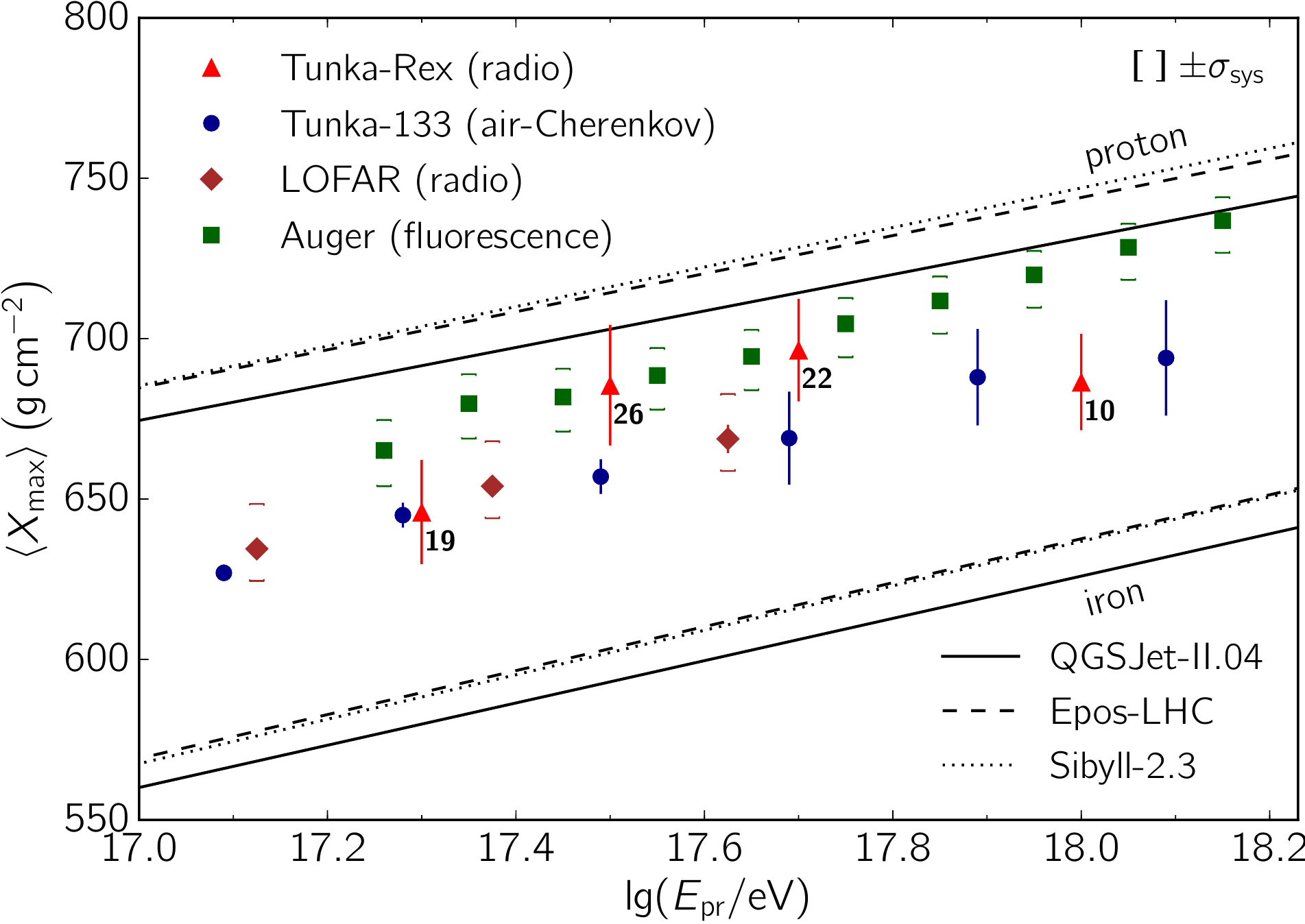}
\end{center}

\end{minipage}\hspace{2pc}%
\begin{minipage}{17pc}
\begin{center}
\includegraphics[height=13pc]{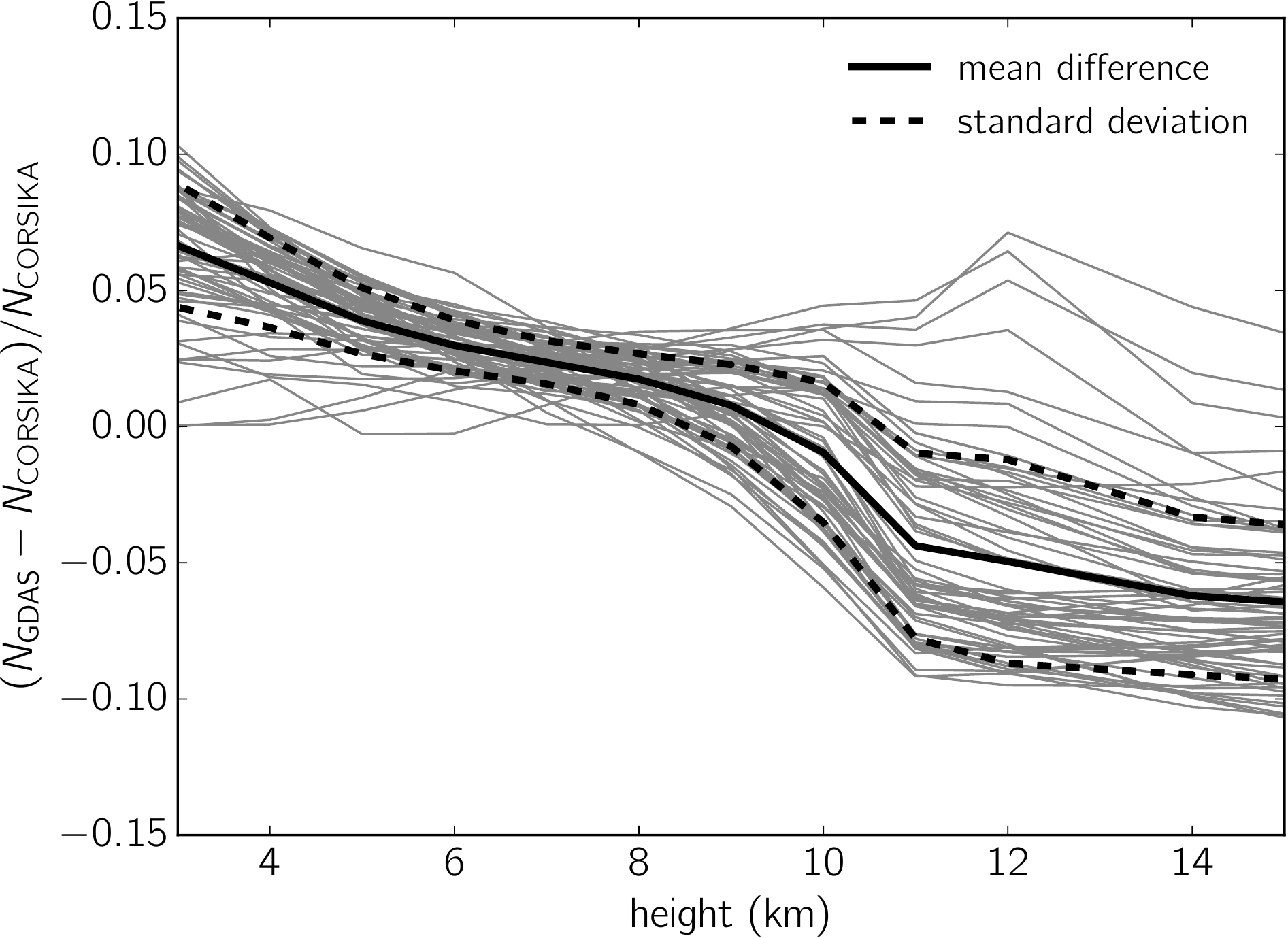}
\end{center}

\end{minipage}
\caption{\label{xmax}{\it Left}: Mean shower maximum as function of primary energy reconstructed by different experiments
measuring the electromagnetic component of air showers~\cite{trex_template}.
        {\it Right}: difference between the refractivity at the CORSIKA~\cite{corsika} standard atmosphere and the one calculated with GDAS profiles
        for the Tunka location.}
\end{center}
\end{figure}

Results of applying this method to the data are shown in the Fig.~\ref{xmax}, left~\cite{trex_template}.

To better understand atmospheric impact
we compared the predictions of the refractivity given by the standard CORSIKA atmosphere
and the values given by GDAS~\cite{gdas}.
The results of this comparison are displayed in the Fig.~\ref{xmax}, right.

Also, mainly due to the worsening of the radio noise environment on the site
some important updates have been introduced in the signal processing chain~\cite{kostunin_arena18}:
\begin{itemize}
  \item Matched filtering and background suppression with convolutional neural networks~\cite{shipilov_arena18, marshalkina_arena18}.
  \item A sliding window for noise estimation was introduced to make the noise estimation procedure more robust against RFI.
\end{itemize}

Tunka-Rex has resolution of the X$_{max}$ reconstruction about 30~g/cm$^2$.
The resolution of the energy reconstruction is approximately 10\%
with systematic uncertainties of the instrument absolute calibration of about 20\%.

\section{Aperture and Exposure Estimation}

The estimation of the instrument exposure
is a key component for the correct reconstruction of
the cosmic-ray flux measured by the observatory.

The general way to estimate the flux $J$ as a function of energy of the primary particle $E$ is
the following~\cite{auger_exposure}:
\begin{equation}
  J(E) = \frac{\mathrm{d}^4 N}{\mathrm{d} E \, \mathrm{d} s \, \mathrm{d} \Omega \, \mathrm{d} t} \approx \frac{\Delta N_{sel}(E)}{\Delta E} \frac{1}{\varepsilon(E)},
\end{equation}
where $N$ is the number of primaries with energy between $E$ and $E+\mathrm{d} E$
passing through a surface element $\mathrm{d} s$ within a solid angle $\mathrm{d}\Omega$ and within time $\mathrm{d}t$.
$\Delta N_{sel}(E)$ is the number of events passing the selection criteria in a given analysis
in the energy bin centered in $E$ and having a width $\Delta E$.
$\varepsilon(E)$ is the energy-dependent exposure.

The exposure can be expressed in the form of the integral of the detection
efficiency $\xi$:
\begin{equation}
  \varepsilon(E) = \int \limits _{T_{sel}} \int \limits _{\Omega_{sel}} \int \limits _{S_{sel}} \xi \left( E,t,\vartheta,\varphi,x,y \right) \cos \vartheta \, \mathrm{d} s \, \mathrm{d} \Omega \, \mathrm{d} t =
  \int \limits _T A (E,t) \, \mathrm{d} t,
  \label{exposure}
\end{equation}
with $\mathrm{d} \Omega = \sin \vartheta \, \mathrm{d} \vartheta \, \mathrm{d} \varphi$ and $\mathrm{d} s = \mathrm{d} x \,\mathrm{d} y$.
After partial integration one obtains the value $A$, which is called instant aperture.
The detection efficiency depends strongly on energy, direction, and shower core location
as well as on observation time due to instrumental instability over time.

The idea is to use of the full efficiency regime
and fiducial selection of area and angular coverage of the sky.
Using this regime of full efficiency
is the most reliable way to estimate the cosmic-ray flux since
instrumental effects of changing the observed mass composition near the detection threshold
do not influence the final results.
In this case the aperture in the equation~(\ref{exposure}) can be factorized
in the following way
\begin{equation}
  \varepsilon(E) = \int \limits _T \, S (E,t) A_{\Omega} (E,t) \, \mathrm{d} t,
  \label{full-efficiency}
\end{equation}
where $S (E,t)$, $A_{\Omega} (E,t)$ are the fiducial area of the instrument and
the angular part of the aperture with fiducial selection of the viewing angle.
Full efficiency means that the detection efficiency is equal to one,
and is omitted hereafter.
In the simplest case of stable operation the equation~(\ref{full-efficiency})
can be simplified
\begin{equation}
  \varepsilon(E) = S(E) A_{\Omega}(E) T,
\end{equation}
where $T$ is the observation time at full efficiency.

In comparison to particle and non-imaging air-Cherenkov detectors,
the full efficiency regime of radio measurements in the bandwidth we use
has a distinguishing feature:
the detection efficiency for arrival directions near the geomagnetic field is suppressed.
This makes straightforward calculations of the instrumental full-efficiency angular coverage
impossible.

Recently we developed a model taking into account this feature.
This model uses elliptical shower footprints
for calculation of detection efficiency for a given area and incoming direction.
Parametrization of the ellipse parameters in this model is obtained from our experimental data.

This simple model, however, does not describe the observed detection efficiency in required detail.
It works sufficiently well only in the intermediate range of zenith angles.
Comparisons of the model against the hybrid observations with TAIGA detectors
as well as detailed description of the model can be found in~\cite{trex_exposure}.

Calculations of aperture can be done with high precision by utilizing an analytical approach.

Since the main problem in the analytical calculation of the angular part of the aperture is
the suppression around the magnetic field direction we can introduce a simple cut around this region.
An additional cut for low zenith angles
is introduced
due to the low number of triggered antennas in this region.

Utilizing the considered cuts one obtains:
\begin{equation}
  A_{\Omega} (E) = 2 \int \limits _{\vartheta_{min}\left(E\right)} ^{\vartheta_{max}}
    \left[ \pi
    - \arccos \left( \frac{\cos \alpha_{min}\left(E\right) - \cos \vartheta \cos \vartheta_m}{\sin \vartheta \sin \vartheta_m} \right)
    \right] \cos \vartheta \sin \vartheta \, \mathrm{d} \vartheta.
    \label{a_omega}
\end{equation}
In this equation $\vartheta_{max}$ is the maximal zenith angle,
$\vartheta_{min}$ --- minimal zenith angle,
$\alpha_{min}$ --- minimal geomagnetic angle required
(recall that geomagnetic angle is the angle between the arrival direction and the geomagnetic field),
$\vartheta_{M}$ --- zenith angle of geomagnetic field.
These cuts are applied in the fiducial selection of the events.
The integral is solved numerically.

Details of the derivation this formula can be found in the appendix.

The presented method allows calculation of the experimental aperture semi-analytically
taking advantage of energy dependent cuts.
The choice of cut parametrization for the particular mode is under investigation as well as overall performance of the method.

\section{Conclusion}

Tunka-Rex has been operating and successfully measuring cosmic rays since 2012.
Reliable reconstruction methods and instrument calibration were developed during the last years.
The main results of the Tunka-Rex can be summarized as following:
\begin{itemize}
  \item Development of a robust reconstruction method for data containing only a few antennas with signals~\cite{trex_template, trex_reco}.
  \item Experimental determination of the energy (10\%) and X$_{max}$ (30 g/cm$^2$)
    precision by comparing radio and air-Cherenkov measurements of the same events~\cite{cross-check}.
  \item Comparing the energy scales of the Tunka-133 and KASCADE-Grande instruments with its radio extensions~\cite{energy_scale}.
  \item Development of an aperture and exposure model~\cite{trex_exposure}
    and utilizing this model for the reconstruction of the mean shower maximum as function of energy~\cite{trex_template}.
\end{itemize}

\ack
  The construction of Tunka-Rex was funded by the German Helmholtz association and the Russian Foundation
  for Basic Research (Grant No. HRJRG-303).
  This work has been supported by the Helmholtz Alliance for Astroparticle Physics (HAP),
  the Russian Federation Ministry of Education and Science (Tunka shared core facilities, unique identifier RFMEFI59317X0005, agreements: 3.9678.2017/8.9, 3.904.2017/4.6, 3.6787.2017/7.8, 3.6790.2017/7.8),
  the Russian Foundation for Basic Research (grants 16-02-00738, 17-02-00905, 18-32-00460)
  In preparation of this work we used calculations performed on
  the HPC-cluster “Academician V.M. Matrosov” [36] and
  on the computational resource ForHLR II funded by
  the Ministry of Science, Research and the Arts Baden-W\"urttemberg and DFG (``Deutsche Forschungsgemeinschaft'').
  A part of the data analysis was performed using
  the radio extension of the Offline framework developed
  by the Pierre Auger Collaboration~\cite{offline}.

\appendix

\section{Details on Equation~\ref{a_omega}}

\begin{figure}[h]
  \includegraphics[height=15pc]{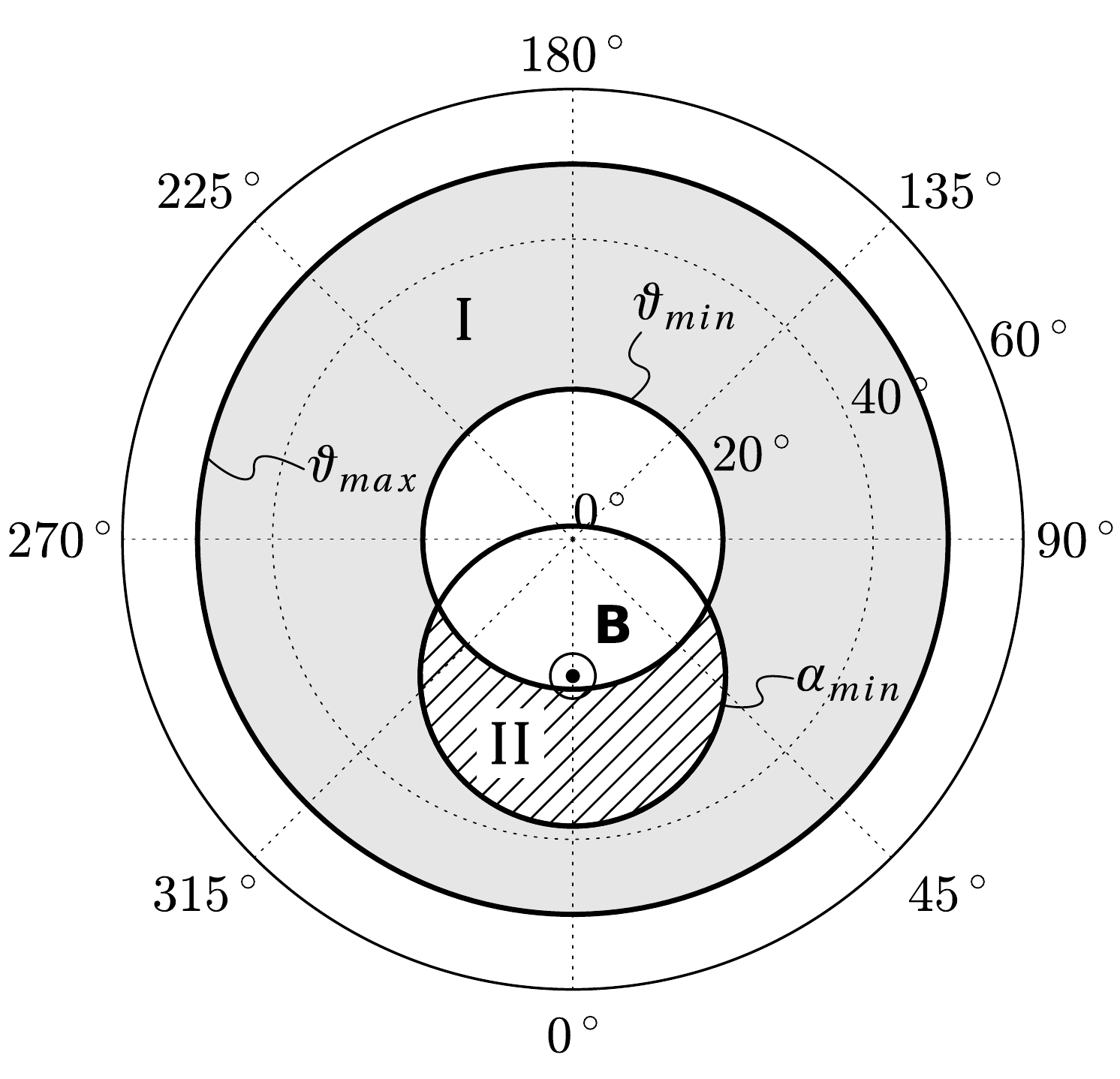}\hspace{2pc}%
  \begin{minipage}[b]{0.5\textwidth}
  \caption{\label{limits}Cuts and areas used in the calculations
    (azimuthal equidistant projection of the sky). The integral is over the gray area.}
  \end{minipage}
  \end{figure}

  The idea of the derivation of equation~\ref{a_omega} can be illustrated with Fig.~\ref{limits}.
  First, we perform the integration of the cosine over area I, and then exclude the integral over area II.
  The coordinate system is chosen for simplification in a way that the azimuthal angle of the center of area II is zero.

  Let us begin with the integral for area I.
  It is useful to recall that cosine in the initial integral holds information that
  the detector is flat.
  \begin{equation}
      (I) = \int \limits _{\Omega'} \cos \vartheta \, \mathrm{d} \Omega =
      \int \limits _{0} ^{2\pi} \int \limits _{\vartheta_{min}} ^{\vartheta_{max}} \cos \vartheta \sin \vartheta \, \mathrm{d} \vartheta \, \mathrm{d} \varphi =
      2\pi \int \limits _{\vartheta_{min}} ^{\vartheta_{max}} \cos \vartheta \sin \vartheta \, \mathrm{d} \vartheta.
  \end{equation}
  This equation can be integrated further
  \begin{equation}
   (I) = \pi \left[ \cos ^2 \vartheta_{min} - \cos ^2 \vartheta_{max} \right].
  \end{equation}
  This is the standard equation for the calculation of the aperture in the case of full efficiency
  and absence of any azimuth-dependent cuts. If $\vartheta_{min}=0$ then
  $(I) = \pi \left[ 1 - \cos ^2 \vartheta_{max} \right].$

  For the area II we begin with the same equation and exploiting the symmetry for the integration:
  \begin{equation} \label{ii}
      (II) = \int \limits _{\Omega'} \cos \vartheta \, \mathrm{d} \Omega =
      2 \int \limits _{\tilde{\Omega}} \cos \vartheta \, \mathrm{d} \Omega =
      2 \int \limits _{\vartheta_{min}} ^{\vartheta_{max}} \int \limits _{0} ^{f(\vartheta)} \cos \vartheta \sin \vartheta \, \mathrm{d} \varphi \, \mathrm{d} \vartheta.
    \end{equation}
  As upper limit of the integral over $\varphi$ we use dependency of $\varphi$ on $\vartheta$
  on the circle of radius $\alpha_{min}$ around the direction to
  the local magnetic field ($\vartheta_M$, $\varphi$).
  We use equation for circle on spherical surface adopted from~\cite{bronshtein}
  \begin{equation}
    \cos \alpha_{min} = \cos \vartheta \cos \vartheta_M + \sin \vartheta \sin \vartheta_M \cos \varphi.
  \end{equation}
  Hence it is easy to express the required dependency
  \begin{equation}
    \varphi = \arccos \left( \frac{\cos \alpha_{min} - \cos \vartheta \cos \vartheta_M}{\sin \vartheta \sin \vartheta_M} \right).
  \end{equation}
  Using this equation as the upper limit in~\ref{ii} the integral is
  \begin{equation}
      (II) = 2 \int \limits _{\vartheta_{min}} ^{\vartheta'_{max}}
      \arccos
      \left( \frac{\cos \alpha_{min} - \cos \vartheta \cos \vartheta_M}{\sin \vartheta \sin \vartheta_M}
      \right)
      \cos \vartheta \sin \vartheta
      \mathrm{d} \vartheta.
    \end{equation}
    Here $\vartheta'_{max}$ is the maximal zenith angle of area II.

  The final step is to perform the subtraction
  \begin{equation}
      (I) - (II) =
       \pi \left[ \cos ^2 \vartheta_{min} - \cos ^2 \vartheta_{max} \right]
       - 2 \int \limits _{\vartheta_{min}} ^{\vartheta'_{max}}
      \arccos
      \left( \frac{\cos \alpha_{min} - \cos \vartheta \cos \vartheta_M}{\sin \vartheta \sin \vartheta_M}
      \right)
      \cos \vartheta \sin \vartheta
      \mathrm{d} \vartheta.
  \end{equation}
This form is convenient for practical calculations.
However, it can be can be expressed in the following shorter form
\begin{equation}
      (I) - (II) = 2 \int \limits _{\vartheta_{min}} ^{\vartheta_{max}}
    \left[ \pi
    - \arccos \left( \frac{\cos \alpha_{min} - \cos \vartheta \cos \vartheta_M}{\sin \vartheta \sin \vartheta_M} \right)
    \right] \cos \vartheta \sin \vartheta \, \mathrm{d} \vartheta.
  \end{equation}

\end{document}